\documentclass[10pt, twocolumn]{article}

\usepackage{amsfonts}
\usepackage{graphics}
\newcommand{\mc}[2]{{\mathcal{#1}}_{#2}}

\title{Algorithmic Self-Assembly of DNA Tiles and its Application to Cryptanalysis}

\author{ {\bf Olivier Pelletier$^1$ } \\
$^1$Accenture Technology Labs \\
Sophia Antipolis, France\\
\textbf \small Olivier.Pelletier@accenture.com
\and {\bf Andr\'{e} Weimerskirch$^{2,1}$}  \\
$^2$Electrical Eng. \& Information Sciences Dept.  \\
Ruhr-Universit\"{a}t Bochum \\
Bochum, Germany \\
\textbf \small weika@crypto.ruhr-uni-bochum.de \\}

\date{}

\begin{document}

\maketitle

\begin{abstract}
The early promises of DNA computing to deliver a massively parallel architecture well-suited to computationally hard problems have so far been largely unkept. Indeed, it is probably fair to say that only toy problems have been addressed experimentally. Recent experimental development on algorithmic self-assembly using DNA tiles seem to offer the most promising path toward a potentially useful application of the DNA computing concept. In this paper, we explore new geometries for algorithmic self-assembly, departing from those previously described in the literature. This enables us to carry out mathematical operations like binary multiplication or cyclic convolution product. We then show how to use the latter operation to implement an attack against the well-known public-key crypto system NTRU. 

\end{abstract}

\begin{center}
\tt {Keywords: DNA, self-assembly, multiplication, convolution product, cryptanalysis, NTRU, Wang tiles}
\end{center}

\section{Introduction}
Since the seminal work of Adleman on the Traveling Salesman Problem (TSP) \cite{adleman2}, DNA computing has received a lot of attention both from a theoretical point of view \cite{lipton1} and from an experimental perspective \cite{ouyang1,liu1}. By using the DNA molecule as a carrier of non-genetic information, and biochemistry as a way to process this information, it is possible to build a massively parallel computing architecture. The implementation details vary from one experimental approach to another, but it is certainly fair to describe the overwhelming majority of the reported experiences in the following way: DNA molecules are used to represent potential solutions and biochemical reactions are used to test whether these solutions satisfy or not the criteria for being an actual solution of the problem. Even though a single biochemical step can take as much as one day to perform, the number of solutions tested in parallel is of the order of the Avogadro number (that is $10^{23}$ molecules), opening interesting computational perspectives. Several authors have described how DNA computing could be used to solve difficult problems like boolean satisfiability \cite{lipton1}. Adleman himself has proposed an application to crack the DES encryption scheme \cite{adleman1}. DES is a standard symmetric encryption algorithm with a key of 56 bits, and a few test tubes of DNA would be enough to carry out a brute force attack on this cipher. Unfortunately, these approaches suffer from a number of drawbacks: (1) they are not always easy to implement in biochemistry, especially because they require purification steps; (2) they rely essentially on brute force because they do not easily make use of additional information that might be available about the problem. The price to pay for massive parallelism is a restricted flexibility in "programming" the DNA molecules. For cryptographic applications, this means that the "traditional" approach cannot take advantage of the known attacks on the weaknesses of a given algorithm. 

Mao et al. \cite{mao1} have recently shown that some degree of flexibility can be introduced in DNA computing while retaining the intrinsic advantage of massive parallelism. For that purpose they used DNA tiles that are a biochemical implementation of the mathematical concept of Wang tiles \cite{wang1}. We will describe these objects in more detail in Section 2. Suffice it to say for now that the algorithmic self-assembly of Wang tiles is Turing Universal and that Mao et al.~demonstrated the experimental feasibility of this concept. We feel that it is therefore appropriate to investigate in more detail to what extent the algorithmic self-assembly of DNA tiles can be used to solve problems that could not be practically solved using the "traditional" approach based on the self-assembly of linear DNA. In what follows, we demonstrate in particular that binary multiplication and cyclic convolution product are relatively straightforward to implement (Section 3). Furthermore, we show that the practical implementation of our ideas requires the creation of a finite number of tiles well within the reach of current combinatorial chemistry. Finally we discuss how our ideas could be used to implement a cryptanalytic attack on the well-known public-key crypto system NTRU (Section 4).  

\section{Algorithmic Self-Assembly}

\subsection{Wang Tiles}
The concept of algorithmic self-assembly is closely related to that of Wang tiles. Wang showed that square tiles with colored edges can emulate a Turing machine, if they are allowed to assemble in a way that would cover the plane, according to additional rule that edges of the same color have to face each other \cite{wang1}. This can be intuitively understood by thinking of a given row of tiles as representing a state of the Turing machine while the color encoding plays the role of the matching rules. This shows that computing using Wang tiles is universal~\cite{winfree1,rothemund2}. 

\subsection{Physical Implementation of Wang Tiles}
Recent advances in the field of materials science have enabled the experimental study of algorithmic self-assembly (abbreviated as ASA in the following). The first system, studied by Rothemund \cite{rothemund1}, was made of tiles whose edges were coated with materials of different hydrophobicity. The error rate was found to be unacceptably high, even though some expected distinctive features were observed. We remark that the hydrophobic/hydrophilic interaction used in these experiments was probably not specific enough to enforce proper edge matching (despite some very clever ``ad-hoc'' tricks used by the author). More recently Mao et al.~\cite{mao1} have shown that nanoscopic tiles can be manufactured using DNA that are the molecular equivalent of Wang tiles. These ``Triple Crossover'' tiles are made of several strands of DNA interwoven to create a square body made of DNA double helixes with single (reactive) strands of DNA sticking out from each edge of the tile. In the following we will refer to those single strands of DNA as sticky ends because they have the ability to bind to their Watson-Crick complement. This mechanism corresponds to the color matching rule in the abstract Wang tiles system. The experimental investigation focused on the XOR operation of two binary strings. The size of the problem was still relatively small, but the result turned out to be promising, with an error rate that was less than 2\%. Therefore DNA seems to be the material of choice to implement ASA on a wider scale. Indeed, materials scientists have achieved a high degree of control over the nanostructures that can be built using DNA, and the interaction between single strands of DNA seems to be specific enough to enable a self-assembly with an acceptable error rate. 
 
\subsection{Our Perspective: Algorithmic Self-Assembly for Practical Problems}
Even though ASA is universal, it does by no means follow that any problem can be practically addressed by this approach. Indeed, traditional DNA computing is also universal but, as mentioned above, the quantity of materials needed to perform a calculation prevents it to be used for anything but toy problems. Why is there any reason to believe ASA is a more interesting approach? The answer lies in the fact that DNA tiles can be more easily ``programmed'' to incorporate the constraints of a given problem. It is therefore possible to exercise some degree of control over the biochemical reaction occurring in the test tubes, thus avoiding the considerable waste of materials that characterize the traditional approach. Given the recent experimental developments mentioned above, we believe it is timely to reflect on the best use that could be made of ASA for practical purposes. Our approach is resolutely constructive: we try to provide examples where ASA turns out to be a practical way of solving otherwise difficult problems. This means that we have to depart from the only geometry that has been studied so far (square with four sticky ends). We give examples where a bigger number of sticky ends or a self-assembly not constrained to proceed in a plane turn out to be advantageous. To use a very bold analogy, this is reminiscent of the common situation in traditional computer science where a problem is straightforward to program in a given language (say C) while it is hard to address in another one (say assembly) \footnote{The analogy breaks down pretty quickly as one tries to give it a more formal shape, but we hope it is still useful to carry our message.}.  

\section{Mathematical Operations in DNA}
In this section we describe how to perform mathematical operations in DNA for two examples. First we show how to execute a multiplication in 2D. Then we introduce a method to carry out ASA in three dimensions to execute a cyclic convolution product. We give an abstract overview of each operation, and then go more into details. Note that we did not do any practical experiments. 

\subsection{Multiplication}
\begin{figure}[ht]
\centering
\includegraphics{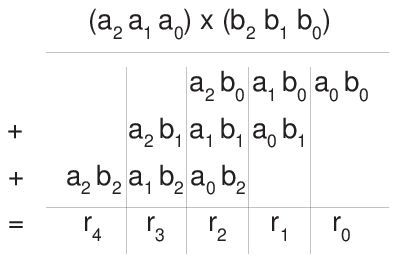}
\caption{Multiplication schoolbook method.}
\label{multbasic}
\end{figure}

We implement the schoolbook method as shown in Figure~\ref{multbasic}. As example we use a multiplication of two 3-bit numbers. The binary input is given as vectors $a$ and $b$ with result $r$ as sum of the corresponding rows under respect of carry overs. The spatial layout of the DNA after self-assembly is very similar to that of electronic circuits carrying out the same function~\cite{basic electronics book}. 

\begin{figure}[ht]
\centering
\includegraphics{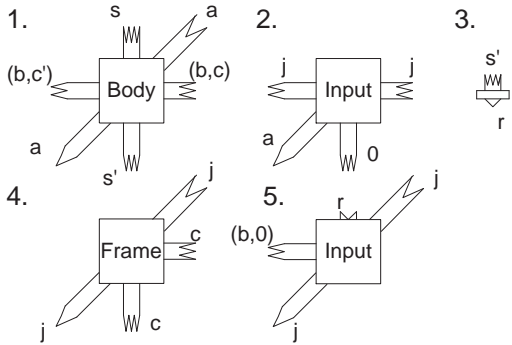}
\caption{DNA tiles for multiplication.}
\label{dnatiles}
\end{figure}

Figure~\ref{dnatiles} depicts the basic DNA tiles that are needed. Tile (1) is used for the actual execution. The original binary operand values are represented by $a$ and $b$ while $s$ and $c$ represent the intermediate sum and carry over, respectively. The result of this elementary operation, intermediate value and carry over, are denoted by $s'$ and $c'$. It follows that
\[ c' = (ab + c + s)/2, \ s' = (ab + c + s) \bmod 2 \]
where integer division is used. There are $16$ different input values determining the number of different tiles of this kind. Tiles (2) and (5) are used to represent operand bits. The connection to the next and previous input tile is denoted by $j$, while the final result of a column is connected at $r$. Tile (3) represents a result bit which will connect to the sticky end $r$ of input tile (5) and the sum $s'$ of tile (1). Furthermore we use frame tiles to limit the physical expansion of the execution. Frame tile (4) forwards the carry over value $c$ to the next left column. Further auxiliary tiles are used (start and end). 

\begin{figure}[ht]
\centering
\includegraphics{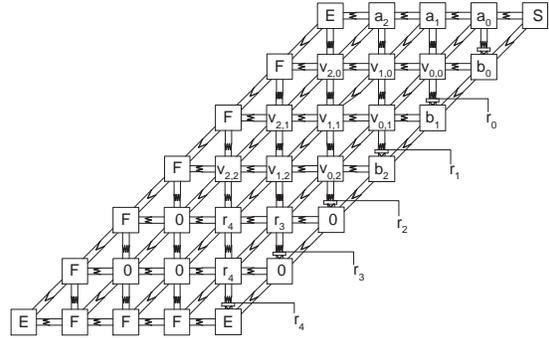}
\caption{Multiplication in DNA.}
\label{mult}
\end{figure}

Figure~\ref{mult} shows the arrangement of the DNA tiles to perform a multiplication. Note that we pad the second operand $b$ with 0-bits to make reading of the result easier, and that the result tile connected to $b_0$ is not part of the result. The body tiles are denoted by $v_{i,j}$, input tiles as $a_i$ and $b_j$ respective, and frame tiles by $F$. Extra tiles are needed as starting and end point, denoted as $S$ and $E$. We understand that different kind of tiles need different sticky ends to avoid ambiguity. However, there are enough combinations available~\cite{feldkamp1}. It is clear that this method can be applied to bigger operands, and that it does not require the operands to have the same length. 

So far we have assumed that linear assemblies of input tiles could be readily obtained. We now outline the way these inputs are ``synthesized''. Given a binary string of $N$ bits, we need $2N$ different tiles indexed by their value (0 or 1) and their position within the string (the DNA sequence connecting one digit to the next one is of course unique for each pair of value and position). Creating a given input simply consists in picking out $N$ such tiles with different indices. Note that, given the appropriate supply, if all the $2N$ tiles are mixed together it is possible to obtain the $2^N$ possible binary strings in a combinatorial fashion. By using non-identical concentrations for the two possible values at a given position, it is also possible to induce a probability distribution on the input strings. All in all, prior to any calculation involving two strings of length $m$ and $n$, we need to synthesize sets of $m+n$ different input tiles\footnote{Note that combinatorial $a$ and $b$ requires $B (m+n)$ tile classes where $B$ is the base of $a$ and $b$, i.e., $2(m+n)$ for binary representation.}, $16$ body tiles, $4+2$ frame tiles, $2$ result tiles, $3$ end tiles, and $1$ starting tile. Once the input strings have been synthesized our scheme requires only one reaction step. All the basic types of tiles are mixed together and the self-assembly can proceed. Reading the final result could be done using the reporter strand technique described in \cite{mao1}. We note that in our case the reporter strand would have to run through the entire 2D lattice. Alternatively, one could imagine that each result tile would have a sticky end running perpendicular to the plane of self-assembly. This would allow the formation of a linear self-assembled structure above this plane, that could be used to produce a reporter strand whose size would scale linearly with the size of the solution\footnote{Note that the first ``dummy'' result tile comes in handy as a PCR primer.}. Thus, even in the worst case, our multiplication scheme requires only two reaction steps and the number of different tiles required is growing linearly with the size of the problem. 

\subsection{Cyclic Convolution Product}
\label{convprod}
\begin{figure}[ht]
\centering
\includegraphics{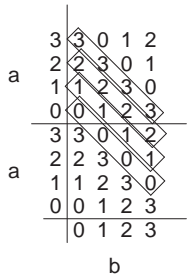}
\caption{Geometrical description of convolution product.}
\label{convbasic}
\end{figure}

After showing an example of ASA using relatively complex tiles to produce a straightforward 2D self-assembly, we now introduce an operation which can be performed more conveniently in 3 dimensions. First we define the cyclic convolution product. Let $F = \sum_{i=0}^{N-1} F_i x^i = [F_0, \ldots, F_{N-1}]$ be a polynomial or a vector of length $N$. Then the cyclic convolution product $\star$ of two vectors of length $N$ is defined as~\cite{hoffstein1}:
\begin{eqnarray*}
A \star B & = & C \ \textrm{with} \\
C_k & = & \sum_{i=0}^{k} A_i B_{k-i} +  \sum_{i=k+1}^{N-1} A_i B_{N+k-i} \\
 & = & \sum_{i+j \equiv k \bmod N} A_i B_j 
\end{eqnarray*}
The $\star$ multiplication modulo $q$ means that the coefficients $C_k$ are reduced by $q$. From now on we will focus on the modulo product. Figure~\ref{convbasic} gives a geometrical description of the convolution product. The input operands are the vectors $a$ and $b$. The $x$ and $y$ axis describe the index of the operand bits. The figure shows the index $k$ of $c_k$ to which $a_i b_j$ contributes. By repeating the input vector $a$ the result coefficient $c_k$ can easily be obtained by adding the diagonal elements. 

\begin{figure}[ht]
\centering
\includegraphics{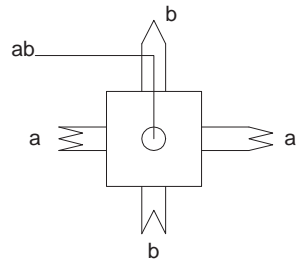}
\caption{Body tile for convolution product.}
\label{convtiles}
\end{figure}

\begin{figure}[ht]
\centering
\includegraphics{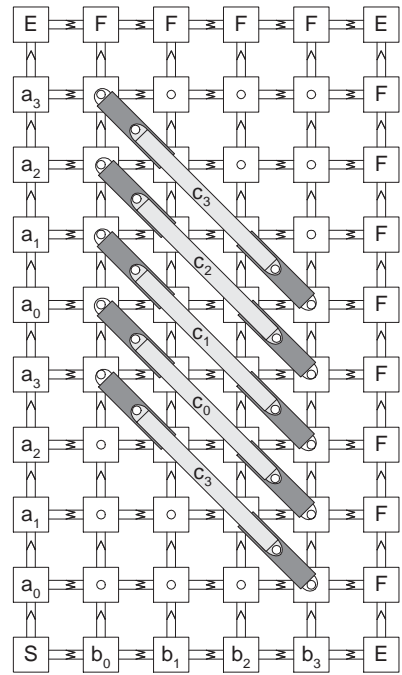}
\caption{Cyclic convolution product in DNA.}
\label{conv}
\end{figure}

We execute the convolution product in DNA according to the geometry just outlined. First we assemble the elementary multiplications in a ``ground layer'', then we grow the crystal to the third dimension to obtain the result. Figure~\ref{convtiles} describes the body tile of the ground layer. It has two input ends $a$ and $b$, forwards the input to the opposite side, and outputs the value $a b$ using a sticky end pointing in the direction perpendicular to the plane of self-assembly \footnote{It is depicted as a circle in Figure~\ref{convtiles}.}. Figure~\ref{conv} shows how the ground layer is built. Again we use input tiles, frame tiles, and start and end tiles. Note that the first operand $a$ is fixed since it has to be repeated.

\begin{figure}[ht]
\centering
\includegraphics{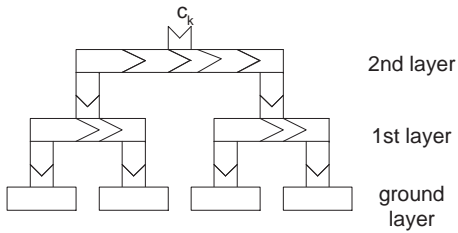}
\caption{Bridge to add the coefficients.}
\label{bridges}
\end{figure}

To add the coefficients, we use bridges which are assembled for each layer beforehand. The implementation of the bridges ensures that the result coefficients are modulo reduced. The bridges are built using connectors to the lower layer, a connector to the next layer, and spacer tiles\footnote{whose number depends on the layer under consideration}. Bridges broken down into their constitutive tiles are shown in Figure~\ref{bridges}. 

\begin{figure}[ht]
\centering
\includegraphics{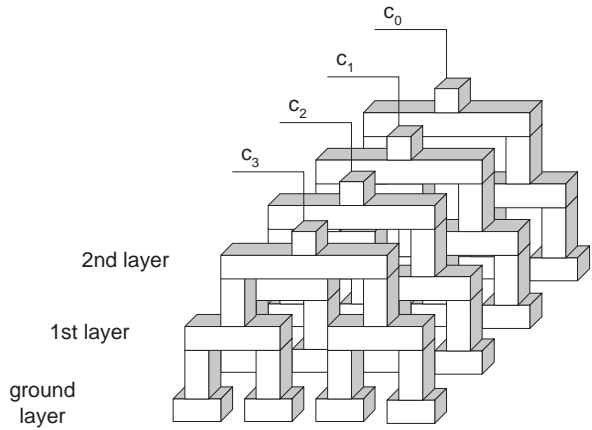}
\caption{Bridges in the 3D space (connections in the ground layer are not pictured for simplification).}
\label{bridges3d}
\end{figure}

The three dimensional arrangement of the bridges is shown in Figure~\ref{bridges3d}. The bridges are arranged on top of the ground layer as outlined in Figure~\ref{conv}. The dark grey connections represent bridge connections in the first layer, while the light grey connections stand for second layer bridges. 

To simplify the bridge building operation we assume that the operands have a length which is a power of 2. To force the bridges to operate along the appropriate diagonal, it would be necessary to use a 2D lattice with a lower symmetry than the square symmetry used for convenience in Figure \ref{conv} \footnote{To prevent that bridges are attached to the wrong tiles such that at some sticky ends there is no bridge attached at all one could use input tiles for $a$ with alternate length in between.}. The result of the operation can now be read at the sticky ends of the uppermost layer. The coefficient $C_{N-1}$ appears twice and has to be ignored once. The diagonals which we do not consider will not assemble up to the highest layer.

Note that input coefficients are integers instead of binaries. Therefore the number of tiles needed is much larger than for the multiplication. Let $a_i \in \{0, \ldots, s-1\}$ and $b_j \in \{0, \ldots, t-1\}$. Synthesis of the input tiles requires $2N + N$ tile classes. Fixed $a$ and combinatorial $b$ requires $2N + tN$ tile classes though. Including start, end, frame, input, and body tiles we need $1 + 3 + 4 + 2N + N + st = 8 + 3N +st$ tile classes for the first layer. Remember that the bridge tiles perform addition modulo $q$. Therefore we need $q^2$ different bridge combinations, i.e., $q^2$ connector tiles to the upper layer and $2q$ connectors to the lower layer. Furthermore we need spacer tiles to build the bridges. Assuming that $q$ is considerably larger than $s$ and $t$ the number of different tiles is in the order of $q^2$ even for combinatorial $b$. If $N = 2^x$ then our structure will consist of $x + 1$ layers including the ground layer. Each layer will be grown one at at time, using bridges with the appropriate spacing. The final result will therefore be obtained in $x+1$ steps. 

\subsection{More Practical Considerations}
The first problem that needs to be addressed is that of the error rate during a computation. Theoretical considerations taking into account the thermodynamics of the system are clearly outside of the scope of this paper, and we will therefore only lead a qualitative discussion. The process of self-assembly within a plane, that we use for our multiplication scheme and as the first step in our cyclic convolution product, is in essence very similar to the construction of Mao et al.~for their XOR product. It is therefore likely that an experimental error rate below $2\%$ could be expected. Much higher error rates could be expected for the building of the successive layers in our convolution product because cooperativity is much lower in this direction (the number of neighbors is much lower which means less constraints). This would probably require the interaction energies between the sticky ends in this direction to be relatively high, in order to give a maximum energetic penalty to possible ``orphan'' sticky ends.
 
Any reader familiar with materials science will probably already have more than a few objections to our claims. Indeed, we must acknowledge that, to date, no DNA tile has been synthesized that could be used to implement our schemes directly. To what extent this will be true in the future is of course absolutely impossible to tell. We will simply refer the reader to the recent work accomplished by the group of Seeman \cite{seeman1} on the creation of DNA-based nanostructures and let him decide for himself how far experimental science is from being able to implement our ideas. We do not believe that our computation schemes alone would be enough to motivate the considerable experimental work required to investigate the 3D ASA of DNA. But we should note that this technique is also very promising for the much more researched problem of protein crystallization. It is therefore not completely utopic to expect experimental progress on that front. Also, even though the total number of different types of tiles to be synthesized is not overwhelming and certainly within the reach of combinatorial chemistry techniques, even for operations on binary numbers of a few hundred bits, it remains to be seen which incentive an experimentalist could have to perform such an experiment. That's why we devote the next section of this article to show that it might be possible to implement an attack on a strong public-key crypto system using our strategy for the cyclic convolution product. 

\section{Application to Cryptography}
In the mid 90's it was shown that DNA computing can be applied to break DES~\cite{boneh1,adleman1}. These methods are based on a brute force attack. Using the parallel nature of DNA computing all possible keys are tested. For symmetric encryption ciphers like DES a brute force attack often is the only practical attack due to limited knowledge. However, for public-key methods brute force attacks are usually far out of computing power range because the key length is chosen according to the best known attack. which requires much less effort than brute force. DNA attacks are limited by the complexity of an attack step and the amount of DNA. In the following we will present the public-key system NTRU and a simple brute force attack in DNA. Then we present the execution of an attack on NTRU which reduces the amount of DNA by the square root. DNA tiles provide the appropriate flexibility to implement both attacks.
 
\subsection{Overview of NTRU}
\subsubsection{Notation}
In this section we will give a brief overview of NTRU. For further details see~\cite{hoffstein1}. The NTRU system is based on a ring $R = \mathbb{Z}/(X^N-1)$ , three integers $(N, p, q)$ and four sets $\mc{L}{f}, \mc{L}{g}, \mc{L}{\phi}, \mc{L}{m}$ of polynomials of degree $N-1$ with integer coefficients. We assume that $gcd(p,q)=1$, and that $q$ is considerably larger than $p$. Elements $F \in R$ are written as a polynomial or vector
\[ F = \sum_{i=0}^{N-1} F_i x^i = [ F_0, F_1, \ldots, F_{N-1} ] \]
Multiplication in $R$ is done using the cyclic convolution product $\star$ as defined in Section~\ref{convprod}. Multiplication modulo $q$ means that the coefficients of the convolution product are reduced modulo $q$.  

\subsubsection{Key Creation}
Assume two entities called Bob and Alice who want to exchange messages over an insecure channel. First Bob chooses elements $f \in \mc{L}{f}$ and $g \in \mc{L}{g}$. For simplicity we assume that $f$ has coefficients in $\{0,1\}$ and that $g$ has coefficients in $\{0, ldots, s-1\}$. The polynomial $f$ is chosen such that it has exactly $d$ coefficients of value $1$ and $N-d$ coefficients of value $0$.  Bob computes $f_q^{-1} \equiv f^{-1} \bmod q$ and $h \equiv f_q^{-1} \star g \bmod q$. Bob's private key is the polynomial $f$ and his public key is $h$.  
\subsubsection{Encryption}
To encrypt a plain text message $m \in \mc{L}{m}$ using Bob's public key $h$, Alice selects a random element $r \in \mc{L}{\phi}$ and computes the cipher text $e \equiv (r \star h + m) \bmod q$.

\subsubsection{Decryption}
To decrypt the cipher text $e$ using the private key $f$, Bob first computes $a \equiv f \star e \bmod q$ where he chooses the coefficients of $a$ in the interval from $-q/2$ to $q/2$. Now Bob recovers the plain text message as $m \equiv (f^{-1} \bmod p) \star a \bmod q$. 

\subsection{Brute Force Attack}
The goal of the attack is given all parameters $(N, p, q)$, the sets $\mc{L}{f}, \mc{L}{g}, \mc{L}{\phi}, \mc{L}{m}$, and a public key $h$ to recover the private key $f$. Let us assume that polynomials in $\mc{L}{g}$ have coefficients in $\{0, \ldots, s-1\}$. As before any $f \in \mc{L}{f}$ has $d$ coefficients of value $1$ and $N-d$ coefficients of value $0$. An attacker can recover the private key by trying all possible $f \in \mc{L}{f}$ and testing if $g' = f \star h \bmod q$ has small entries, i.e., if the coefficients are between $0$ and $s-1$. Similarly, an attacker can try all $g \in \mc{L}{g}$ and test if $f' = g \star h^{-1} \bmod q$ has only coefficients $0$ or $1$. In practice, $\mc{L}{g}$ is smaller than $\mc{L}{f}$, so the security is determined by the number of elements in $\mc{L}{g}$. 

The attack can be implemented in DNA as follows. For all $g \in \mc{L}{g}$ compute the cyclic convolution product $g \star h^{-1} \bmod q$ as explained in Section~\ref{convprod}. Choose $h^{-1}$ as the operand which is repeated. Use the massive parallelism of DNA to compute the convolution product of $h^{-1}$ with all the possible $g$. Reading of the operation result is done by the reporter strand method. Among all the results, the one consisting only of 0 or 1 \footnote{Remember that each digit of the result can take $q$ values.} is the private key. To get it we need to run $q-2$ separation steps. 

This attack does not scale up well. It is limited by $q$ which determines the number of different tiles. A typical value for $q$ is $64$ or $128$~\cite{hoffstein1} which means that more than 4,000 different tile classes are needed. Another restriction is given by the total amount of DNA. The number of DNA tiles to be used in the computation cannot be expected to be much more than the Avogadro number (about $10^{23}$). Therefore this kind of attack is roughly limited to $2^{80}$ different possibilities for $g$. Since the coefficients of $g$ are not limited to binary values there can be $2^{80}$ different possible polynomials of length $64$. Usually the key space is defined as the set of possible keys. In our case we extend the definition such that the key space is the set which is used for an attack, i.e., for NTRU this is usually $\mc{L}{g}$ since it is smaller than $\mc{L}{f}$. The key security is defined as the number of steps, or in our case different inputs, that have to be performed or tried before the key is found using the best known attack. The best known attack is shown in the next section and reduces the effort by a square factor, i.e., the key security is the square root of the number of elements in the key space. Thus we can break an NTRU system having a key security of $2^{40}$ for a proper value $s$. However, a typical key security for NTRU in high security scenarios is about $2^{80}$, i.e., $\mc{L}{g}$ has $2^{160}$ elements. In the next section we give a future perspective how this can be achieved.

\subsection{Meet-in-the-Middle Attack}
The meet-in-the-middle attack reduces the effort to find the private key. Compared to a brute force attack this attack reduces the amount of DNA which is required for a successful attack by the square root, or in other words the key space which can be broken is quadratic in size. We will give a brief overview of the attack. Detailed information is given in~\cite{silverman1}. Remember that the private key $f$ has exactly $d$ ones and $N-d$ zeros. The idea of the attack is to search for $f$ in the form $(f_1, f_2)$ where $f_1$ and $f_2$ each have $d/2$ ones and are $N/2$ in length. Then try all possibilities for $f_1$ and $f_2$ such that $(f_1, f_2) \star h \bmod q$ has coefficients between $0$ and $s-1$. This can be done efficiently as follows. 

Choose at random $N/2$ of the $N$ possible positions in $f$. Assume that $d/2$ of these $N/2$ positions have ones in the actual private key $f$. The probability that this assumption holds is approximately $1$ to $\sqrt{d}$, so the following has to be repeated around $\sqrt{d}$ times before $f$ is found. Now relabel the positions such that the chosen $N/2$ positions determine the vector $f_1$ and the other $N/2$ positions $f_2$. The next step is to enumerate over $f_1$. Usually this takes only $N/2 \choose d/2$ steps but in DNA we probably have to iterate over all $2^{N/2}$ binary vectors. Since $d$ is chosen such that the key space is very large the relative difference is very small. First $(f_1,0) \star h \bmod q$ is computed and put into a bin based on its first $k$ coefficients. If the convolution product has coefficients $F_0, \ldots, F_{k-1}$ then it is put into a bin $(I_{j_0}, \ldots, I_{j_{k-1}})$ where $I_{j_i} \supset F_i$ are integer intervals. The size of the intervals are determined by $k$. Then all possible values for $f_2$ are enumerated. The convolution product $-(0,f_2) \star h \bmod q$ is put into a bin $(J_{j_0}, \ldots, J_{j_{k-1}})$ that is defined by intervals that are slightly larger as the previous ones (each bin is exactly by s-1 larger). Finally the bins are compared. If the assumption about the position of the ones in $f$ in the first step was right then there is a matching pair $f_1$ and $f_2$ such that $f_1$ is in $(I_{j_0}, \ldots, I_{j_{k-1}})$ and $f_2$ in $(J_{j_0}, \ldots, J_{j_{k-1}})$, and the private key can be derived as $f = (f_1, f_2)$. 

In DNA the attack is executed as follows. First choose $N/2$ random positions in $f$. The marked positions are represented by $f_1$ which is assembled using DNA tiles such that the marked positions are chosen combinatorial and the unmarked positions are set to $0$. This can easily be done by encoding the position into the DNA tiles that represent $f_1$ as described before. Execute $(f_1,0) \star h \bmod q$ in parallel for all possible combinations of $f_1$ as explained in Section~\ref{convprod}. The polynomial $h$ is the fixed operand which is repeated. Now construct DNA tiles for $f_2$ in the same manner as before but set marked bits to $0$ and iterate over unmarked bits, and execute $-(0,f_2) \star h \bmod q$. To put a product into a bin we use special DNA tiles with two sticky ends that translate an integer value into the corresponding interval. These tiles are different for the two convolution products in the sense that the interval sizes are different. Tiles for the first convolution product can connect to tiles for the second product with the same intervals, i.e., tiles representing $I_{j_i}$ will be glued to tiles representing $J_{j_i}$. Apply these tiles to the convolution products. Assuming that the mobility of the DNA supra molecular assemblies is not too small, two of them will stick together. If such a tandem structure can be found the original assumption was right, and the private key can be determined by reading the input tiles $f_1$ and $f_2$. The actual reading stage is problematic here, as the reporter strand method likely does not seem to work. We suggest another approach where the DNA structures are first filtered according to their molecular weight, and those corresponding to tandem units are examined by atomic force microscopy\footnote{If the recent developments in coupled NMR and AFM become mainstream, then reading could be done by coordinating atoms with very different resonance frequencies to the input tiles.}. 

Assuming that a brute force attack can be mounted to break a key security of $2^{40}$ the described meet-in-the-middle attack in DNA might break systems with a key security of $2^{80}$. However, many assumptions are very optimistic for the near future. Furthermore we understand that using a higher security level, e.g., a key security of $2^{285}$ as proposed in~\cite{hoffstein1} puts public-key systems like NTRU far out of range for a successful cryptanalysis in DNA.  

\section{Conclusions}

We have presented two computation schemes for the binary multiplication and cyclic convolution product using the algorithmic self-assembly of DNA tiles. For that purpose, we introduced new conceptual designs for DNA tiles that should allow for a practical implementation of these operations. Indeed, we emphasize the fact that even though DNA tiles are by themselves universal, tiles with different designs will perform very differently on a given problem: designing effective DNA tiles for a given computation can be thought as ``DNA programming''. The most interesting feature of our system of DNA tiles is that it turns out to be flexible enough to go beyond a simple brute force algorithm: it would indeed be possible to use it to implement an attack on a public-key crypto system. Among the open questions, we have to acknowledge that we do not have any estimations on the expected error rate. We are currently trying to address this issue.

\end{document}